\documentclass[aps,prd,twocolumn,showpacs,superscriptaddress,groupedaddress]{revtex4}  

\usepackage{graphicx}  
\usepackage{dcolumn}   
\usepackage{bm}        
\usepackage{amssymb}   
\def\pbar{$\overline p$}
\def\ol{\overline}
\newcommand{\be}{\begin{equation}}
\newcommand{\ee}{\end{equation}}

\begin{document}

\hspace{5.2in} \mbox{Fermilab-Pub-12-263-E}

\title{Measurement of the differential cross section
$\mathbf{d\boldsymbol{\sigma}/dt}$  in elastic $\mathbf{p\overline{p}}$ scattering  at $\mathbf{\sqrt{s}=1.96}$ TeV\\}

\affiliation{LAFEX, Centro Brasileiro de Pesquisas F\'{i}sicas, Rio de Janeiro, Brazil}
\affiliation{Universidade do Estado do Rio de Janeiro, Rio de Janeiro, Brazil}
\affiliation{Universidade Federal do ABC, Santo Andr\'e, Brazil}
\affiliation{Universidade Estadual Paulista, S\~ao Paulo, Brazil}
\affiliation{University of Science and Technology of China, Hefei, People's Republic of China}
\affiliation{Universidad de los Andes, Bogot\'a, Colombia}
\affiliation{Charles University, Faculty of Mathematics and Physics, Center for Particle Physics, Prague, Czech Republic}
\affiliation{Czech Technical University in Prague, Prague, Czech Republic}
\affiliation{Center for Particle Physics, Institute of Physics, Academy of Sciences of the Czech Republic, Prague, Czech Republic}
\affiliation{Universidad San Francisco de Quito, Quito, Ecuador}
\affiliation{LPC, Universit\'e Blaise Pascal, CNRS/IN2P3, Clermont, France}
\affiliation{LPSC, Universit\'e Joseph Fourier Grenoble 1, CNRS/IN2P3, Institut National Polytechnique de Grenoble, Grenoble, France}
\affiliation{CPPM, Aix-Marseille Universit\'e, CNRS/IN2P3, Marseille, France}
\affiliation{LAL, Universit\'e Paris-Sud, CNRS/IN2P3, Orsay, France}
\affiliation{LPNHE, Universit\'es Paris VI and VII, CNRS/IN2P3, Paris, France}
\affiliation{CEA, Irfu, SPP, Saclay, France}
\affiliation{IPHC, Universit\'e de Strasbourg, CNRS/IN2P3, Strasbourg, France}
\affiliation{IPNL, Universit\'e Lyon 1, CNRS/IN2P3, Villeurbanne, France and Universit\'e de Lyon, Lyon, France}
\affiliation{III. Physikalisches Institut A, RWTH Aachen University, Aachen, Germany}
\affiliation{Physikalisches Institut, Universit\"at Freiburg, Freiburg, Germany}
\affiliation{II. Physikalisches Institut, Georg-August-Universit\"at G\"ottingen, G\"ottingen, Germany}
\affiliation{Institut f\"ur Physik, Universit\"at Mainz, Mainz, Germany}
\affiliation{Ludwig-Maximilians-Universit\"at M\"unchen, M\"unchen, Germany}
\affiliation{Fachbereich Physik, Bergische Universit\"at Wuppertal, Wuppertal, Germany}
\affiliation{Panjab University, Chandigarh, India}
\affiliation{Delhi University, Delhi, India}
\affiliation{Tata Institute of Fundamental Research, Mumbai, India}
\affiliation{University College Dublin, Dublin, Ireland}
\affiliation{Korea Detector Laboratory, Korea University, Seoul, Korea}
\affiliation{CINVESTAV, Mexico City, Mexico}
\affiliation{Nikhef, Science Park, Amsterdam, the Netherlands}
\affiliation{Radboud University Nijmegen, Nijmegen, the Netherlands}
\affiliation{Joint Institute for Nuclear Research, Dubna, Russia}
\affiliation{Institute for Theoretical and Experimental Physics, Moscow, Russia}
\affiliation{Moscow State University, Moscow, Russia}
\affiliation{Institute for High Energy Physics, Protvino, Russia}
\affiliation{Petersburg Nuclear Physics Institute, St. Petersburg, Russia}
\affiliation{Instituci\'{o} Catalana de Recerca i Estudis Avan\c{c}ats (ICREA) and Institut de F\'{i}sica d'Altes Energies (IFAE), Barcelona, Spain}
\affiliation{Uppsala University, Uppsala, Sweden}
\affiliation{Lancaster University, Lancaster LA1 4YB, United Kingdom}
\affiliation{Imperial College London, London SW7 2AZ, United Kingdom}
\affiliation{The University of Manchester, Manchester M13 9PL, United Kingdom}
\affiliation{University of Arizona, Tucson, Arizona 85721, USA}
\affiliation{University of California Riverside, Riverside, California 92521, USA}
\affiliation{Florida State University, Tallahassee, Florida 32306, USA}
\affiliation{Fermi National Accelerator Laboratory, Batavia, Illinois 60510, USA}
\affiliation{University of Illinois at Chicago, Chicago, Illinois 60607, USA}
\affiliation{Northern Illinois University, DeKalb, Illinois 60115, USA}
\affiliation{Northwestern University, Evanston, Illinois 60208, USA}
\affiliation{Indiana University, Bloomington, Indiana 47405, USA}
\affiliation{Purdue University Calumet, Hammond, Indiana 46323, USA}
\affiliation{University of Notre Dame, Notre Dame, Indiana 46556, USA}
\affiliation{Iowa State University, Ames, Iowa 50011, USA}
\affiliation{University of Kansas, Lawrence, Kansas 66045, USA}
\affiliation{Kansas State University, Manhattan, Kansas 66506, USA}
\affiliation{Louisiana Tech University, Ruston, Louisiana 71272, USA}
\affiliation{Boston University, Boston, Massachusetts 02215, USA}
\affiliation{Northeastern University, Boston, Massachusetts 02115, USA}
\affiliation{University of Michigan, Ann Arbor, Michigan 48109, USA}
\affiliation{Michigan State University, East Lansing, Michigan 48824, USA}
\affiliation{University of Mississippi, University, Mississippi 38677, USA}
\affiliation{University of Nebraska, Lincoln, Nebraska 68588, USA}
\affiliation{Rutgers University, Piscataway, New Jersey 08855, USA}
\affiliation{Princeton University, Princeton, New Jersey 08544, USA}
\affiliation{State University of New York, Buffalo, New York 14260, USA}
\affiliation{Columbia University, New York, New York 10027, USA}
\affiliation{University of Rochester, Rochester, New York 14627, USA}
\affiliation{State University of New York, Stony Brook, New York 11794, USA}
\affiliation{Brookhaven National Laboratory, Upton, New York 11973, USA}
\affiliation{Langston University, Langston, Oklahoma 73050, USA}
\affiliation{University of Oklahoma, Norman, Oklahoma 73019, USA}
\affiliation{Oklahoma State University, Stillwater, Oklahoma 74078, USA}
\affiliation{Brown University, Providence, Rhode Island 02912, USA}
\affiliation{University of Texas, Arlington, Texas 76019, USA}
\affiliation{Southern Methodist University, Dallas, Texas 75275, USA}
\affiliation{Rice University, Houston, Texas 77005, USA}
\affiliation{University of Virginia, Charlottesville, Virginia 22901, USA}
\affiliation{University of Washington, Seattle, Washington 98195, USA}
\author{V.M.~Abazov} \affiliation{Joint Institute for Nuclear Research, Dubna, Russia}
\author{B.~Abbott} \affiliation{University of Oklahoma, Norman, Oklahoma 73019, USA}
\author{B.S.~Acharya} \affiliation{Tata Institute of Fundamental Research, Mumbai, India}
\author{M.~Adams} \affiliation{University of Illinois at Chicago, Chicago, Illinois 60607, USA}
\author{T.~Adams} \affiliation{Florida State University, Tallahassee, Florida 32306, USA}
\author{G.D.~Alexeev} \affiliation{Joint Institute for Nuclear Research, Dubna, Russia}
\author{G.~Alkhazov} \affiliation{Petersburg Nuclear Physics Institute, St. Petersburg, Russia}
\author{A.~Alton$^{a}$} \affiliation{University of Michigan, Ann Arbor, Michigan 48109, USA}
\author{G.~Alverson} \affiliation{Northeastern University, Boston, Massachusetts 02115, USA}
\author{G.A.~Alves} \affiliation{LAFEX, Centro Brasileiro de Pesquisas F\'{i}sicas, Rio de Janeiro, Brazil}
% ...
%
%
\collaboration{The D0 Collaboration\footnote{with visitors from
$^{a}$Augustana College, Sioux Falls, SD, USA,
$^{b}$The University of Liverpool, Liverpool, UK,
$^{c}$UPIITA-IPN, Mexico City, Mexico,
$^{d}$DESY, Hamburg, Germany,
$^{e}$SLAC, Menlo Park, CA, USA,
$^{f}$University College London, London, UK,
$^{g}$Centro de Investigacion en Computacion - IPN, Mexico City, Mexico,
$^{h}$Universidad Nacional de Asuncion, Facultad de Ingenieria, Asuncion, Paraguay
$^{i}$ECFM, Universidad Autonoma de Sinaloa, Culiac\'an, Mexico
and
$^{j}$Ohio State University, Columbus, OH, USA
$^{\ddag}$Deceased.
}} \noaffiliation
\vskip 0.25cm

\date{June 4, 2012}

\begin{abstract}

We present a measurement of the  elastic
differential cross section $d\sigma(p\ol p \rightarrow p \ol p)$/$dt$  as a
function of the four-momentum-transfer squared $t$.
The data sample corresponds to an integrated luminosity of  \mbox{$\approx$ 31 nb$^{-1}$} collected with the D0 detector using dedicated Tevatron $p \ol p $ Collider operating conditions  at $\sqrt{s} = 1.96\  \mathrm{TeV}$ and covers the range \mbox{0.26 $<|t|<$ 1.2 GeV$^2$}. 
For \mbox{$|t|<$ 0.6 GeV$^2$}, $d\sigma$/$dt$ is described by an exponential function of the form $Ae^{-b|t|}$ with a slope parameter
 \mbox{$ b = 16.86$ $\pm$ $0.10$ (stat) $\pm$ $0.20$ (syst) GeV$^{-2}$}. A change in slope is observed at \mbox{$|t|\approx $ 0.6 GeV$^2$}, followed by a more gradual $|t|$ dependence with increasing values of $|t|$.
\end{abstract}

\maketitle 

\section{\label{sec:intro}Introduction}


 The differential cross section,
$d\sigma$/$dt$, for $p \ol p \rightarrow p \ol p$, where
$t$ is the four-momentum-transfer squared~\cite{tt}, contains information
about proton structure and non-perturbative aspects of
$p$\pbar\ interactions. In the $|t|$ range studied here, the nuclear scattering
amplitude is expected to dominate~\cite{blockcahn}, and with increasing $|t|$,  $d\sigma$/$dt$
 is expected to fall exponentially, followed by a local diffractive minimum, after which $d\sigma$/$dt$ continues
to decrease~\cite{Giacomelli}. 

Studies of $d\sigma$/$dt$ at different center-of-mass energies, $\sqrt{s}$, have
demonstrated an effect known as shrinkage, namely that the slope of the exponential fall-off of the differential cross section as a function of $|t|$ becomes larger with increasing $\sqrt{s}$, and
the $|t|$ value at which the local diffraction minimum occurs is reduced~\cite{Alberi}. It has  also been observed that the shape of the local diffractive minimum is different between $p$$p$ and $p$\pbar\ elastic scattering~\cite{isr:dsdt}.
The elastic differential cross section plays an important role in constraining 
soft diffractive models~\cite{Nagy,Barone,Landshoff} which cannot be directly calculated by perturbative
QCD~\cite{Forshaw}.

In this Article, we present a measurement of the $p$\pbar\  elastic 
differential cross section at $\sqrt{s} = 1.96\  \mathrm{TeV}$ 
in the range \mbox{0.26 $<|t|<$ 1.2 GeV$^2$}, measured using the forward proton detector (FPD)
spectrometer system of the D0 experiment~\cite{FPD}.
Since $p$ and $\bar p$ elastic scattering angles are typically very small (on the order of  
milliradians), they are 
not covered by the main D0 detector. The
elastically scattered protons and antiprotons  are 
tagged with  detectors  inserted in the
beam pipe on either side of the interaction point (IP).
Our measurement extends the  $|t|$ range 
 previously studied by the CDF (\mbox{0.025 $<|t|<$ 0.29 GeV$^2$})~\cite{CDF:B} and by the E710 (\mbox{0.034 $<|t|<$ 0.65 GeV$^2$})~\cite{E710:larget} Collaborations at the Tevatron and constitutes
the first confirmation of a change in the $|t|$ dependence 
of  $d\sigma (p \ol p \rightarrow p \ol p)$/$dt$  at  center-of-mass 
energy $\sqrt{s} = 1.96\  \mathrm{TeV}$. A similar measurement 
in $pp$ collisions at $\sqrt{s} = 7.0\  \mathrm{TeV}$ has recently 
been reported by the TOTEM Collaboration~\cite{totem} showing similar
 trends for the slope of the differential cross section and for the 
position of the local diffraction minimum, although the local minimum 
found in $pp$ elastic scattering is much more pronounced than 
the kink we observe.

This article is organized as follows. First we describe, 
in Sec.~\ref{sec:detector}, the D0 detector, with particular 
emphasis on the FPD. Next, in Sec.~\ref{sec:selection}, we 
discuss various aspects of the selection of the sample of candidate 
elastic scattering events, including the details on the position 
measurements in the FPD, their alignment, the reconstruction 
of $p$ and \pbar\ scattering angles, the background subtraction, the measurements of acceptance and efficiencies. Systematic uncertainties are 
discussed in Sec.~\ref{sec:sys} and the results of this measurement are described in Sec.~\ref{sec:res}.

\section{D0 Detector}
\label{sec:detector}
We briefly describe the elements of the detector that are relevant for the measurement reported here. A detailed description of the D0 detector can be found in reference~\cite{FPD}.
The central tracking system of the D0 detector comprises a 
silicon microstrip tracker (SMT) and a central fiber tracker (CFT), 
surrounded by a 2~T superconducting solenoidal 
magnet. The pseudorapidity~\cite{eta} coverage for the tracking detectors is $|\eta|<3$ for the SMT and $|\eta|<2.5$ for the CFT.
Outside of the superconducting magnet, the liquid-argon and uranium calorimeter is composed of three sections housed in separate cryostats: a
central calorimeter  section covering the pseudorapidity range  $|\eta| < 1.1$ and two end calorimeter  sections  that extend coverage 
to $|\eta|\approx 4.2$~\cite{run1det}.  A muon system~\cite{run2muon}, outside of the calorimetry, consists of  
three layers of tracking detectors and scintillation trigger counters, and  large toroidal magnets for muon momentum measurement. The luminosity monitor (LM) consists of  
plastic scintillator arrays  located at $z = \pm$140 cm (where $z$ is measured from the IP along the nominal direction of the proton beam), and covers the 
pseudorapidity range  $2.7<|\eta|<4.4$. The LM is used to detect non-diffractive inelastic 
collisions and to make an accurate determination 
of the luminosity. 

\subsection{Forward Proton Detectors}
\label{FPDdet}

\begin{figure*}
\centerline{\includegraphics[width=14.0cm,height=5.0cm]{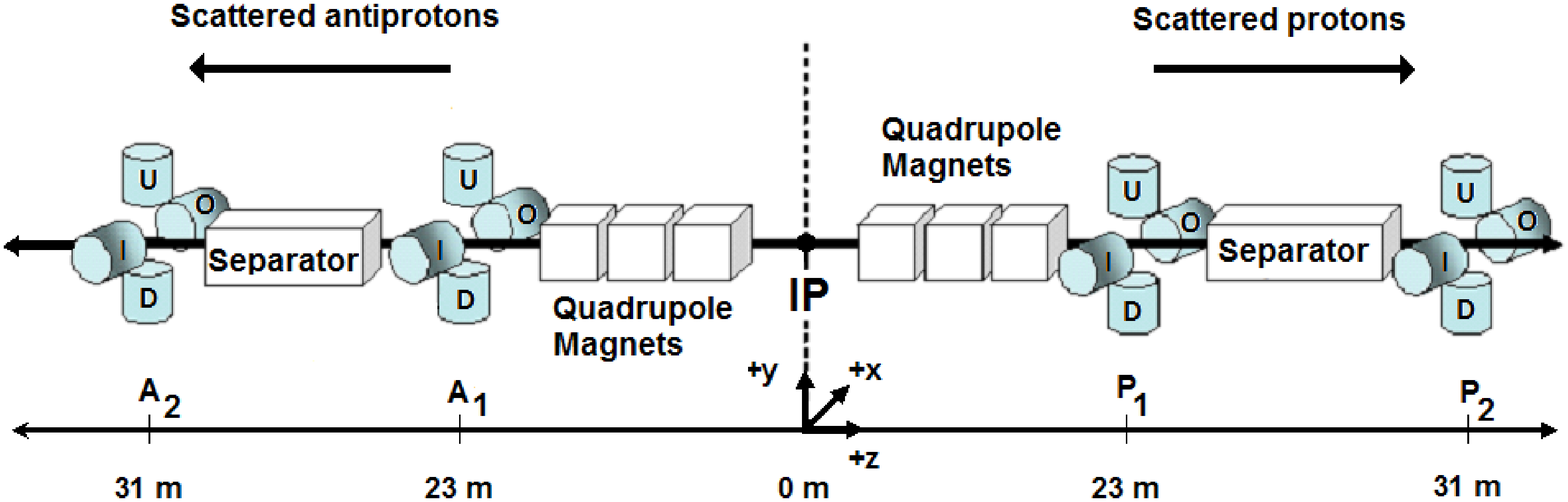}}
\vspace*{0.0in} \caption{Schematic view of the Roman pot stations (A$_1$, A$_2$, P$_1$, P$_2$) comprising the forward proton detector as
described in the text (not drawn to scale). The dipole spectrometer is not shown. The letters U, D, I, O make reference to the up, down, inner and outer detectors, respectively.  } \label{layout}
\vspace*{0.0in}
\end{figure*}

Figure~\ref{layout} shows the layout of the main components of the FPD system relevant to this measurement~\cite{FPD}. In the center of
the diagram is the interaction point, IP, at the center of the main D0 detector. A scattered proton/antiproton goes through 3 Tevatron quadrupole magnets (with a field gradient of about 20 T/m) which alternate defocusing in the horizontal and vertical planes, passes through the first station of detectors, goes through a region free from magnetic field where electrostatic separators are located, and arrives at a second station of detectors. 

The FPD consists of four quadrupole spectrometers on both 
the scattered proton (P) and scattered antiproton (A) sides plus a dipole spectrometer (not shown). Each quadrupole spectrometer is composed of two scintillating fiber detectors, one located at about 23 m (A$_1$ or P$_1$) and the second  at about 31 m (A$_2$ or P$_2$) from the IP along the Tevatron beam line.  Both detectors  are either above (U), below (D), on the
inner side (I), or on the outer side (O) of the beam line. Given the location of the two detectors, the spectrometers on the A side are  named  A$_{\text{U}}$, A$_{\text{D}}$, A$_{\text{I}}$, and A$_{\text{O}}$ (a similar definition is made for the P side). The pseudorapidity range covered by the detectors is about $7.3<|\eta|<8.6$.

\begin{figure}
\includegraphics[scale=0.6]{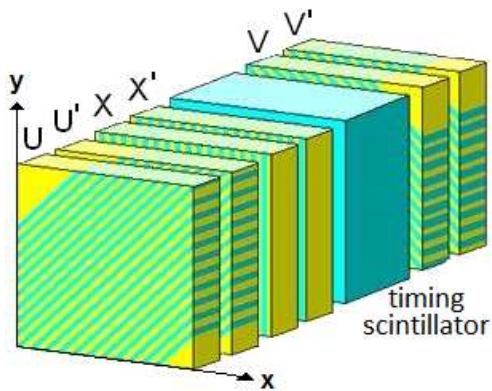}
\caption{Schematic view of the design of one FPD scintillating fiber detector. U, U$^{\prime}$, V, V$^{\prime}$, X, and X$^{\prime}$ are 
the scintillating fiber layers. The local detector $x$, $y$ coordinates are indicated by the arrows. The scintillating fibers are indicated by the green stripes.}
\label{plt:fplanes}
\end{figure}

Scattered protons and antiprotons cross thin stainless steel
windows at the entrance and exit of a vessel 
(Roman pot) containing the detectors~\cite{pot}. The pots are remotely
controlled and moved to within a few millimeters of the beam during stable
 beam conditions.
The Roman pots house position detectors using $0.8$ mm $\times$ $0.8$ mm
square double-clad polystyrene scintillating fibers to detect the
passage of charged particles. Each position detector  consists of six layers 
of scintillating
fibers (U, U$^{\prime}$, V, V$^{\prime}$, X, X$^{\prime}$), where the 
scintillating fibers of the  U, U$^{\prime}$ 
(V, V$^{\prime}$) layers are rotated by plus (minus) 45 degrees with respect 
to the vertical X, X$^{\prime}$ fibers (see Fig.~\ref{plt:fplanes}). 

\begin{figure}
\includegraphics[scale=0.45]{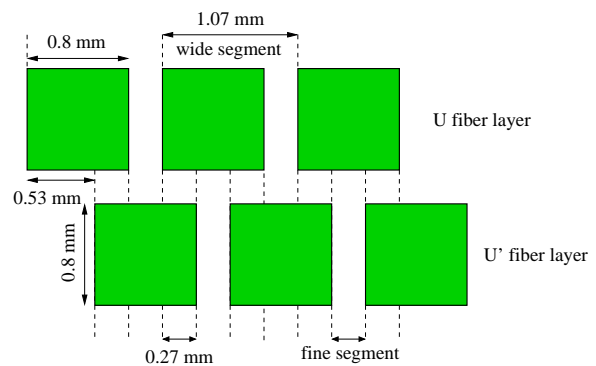}
\caption{Schematic view of part of the U and U$^{\prime}$ fiber layers with the definitions of wide and fine fiber segments. Similar definitions are used for the X, X$^{\prime}$ and V, V$^{\prime}$ fiber layers.}
\label{plt:fiber_offset}
\end{figure}

The fibers of  the ``primed'' planes are offset by  0.53 mm (two-thirds 
of a fiber width) with respect to 
the fibers of the ``unprimed'' layers. 
By combining the fiber information from  ``primed'' and ``unprimed'' layers, we
 obtain ``wide'' fiber segments (about 1.07 mm wide)
 that are used for triggering and  ``fine'' fiber segments (about 0.27 
mm wide) used for offline hit reconstruction (see Fig.~\ref{plt:fiber_offset}).
Each detector also contains a scintillator (read out by a 
 time-to-digital converter system~\cite{FPD}) which provides a time measurement
 for particles passing through the detector with a resolution of about 1 ns. The time measurement  is used to 
distinguish  particles coming from the center of the D0 detector from 
background beam halo particles (particles 
traveling far enough outside of the main beam core that they  pass through the FPD detector).

Beyond the active areas of the detectors, 
matching square clear fibers transport signals from the scintillators  to
 16-channel multi-anode photomultiplier tubes. The electronic signals are subsequently
amplified, shaped, and sent to the D0
triggering and data acquisition system~\cite{FPD}. 

The elastic triggers are defined by the logical OR of coincident hits 
in all four detectors, in either of the four possible configurations 
of a \pbar\ spectrometer and a diagonally opposite (collinear) $p$ 
spectrometer:
A$_{\text{U}}$P$_{\text{D}}$, A$_{\text{D}}$P$_{\text{U}}$, 
A$_{\text{I}}$P$_{\text{O}}$, and A$_{\text{O}}$P$_{\text{I}}$. Several 
different conditions on the hits in the scintillating fiber 
detectors were used in the triggers: a tight (T) trigger
that registered a single hit formed by the coincidence of UU$^{\prime}$, VV$^{\prime}$, XX$^{\prime}$ wide segments; a medium (M)
trigger that allowed up to three wide segments; and a loose (L) trigger that
allowed hits formed from coincidences of two out of the three 
UU$^{\prime}$, VV$^{\prime}$, XX$^{\prime}$ wide segments with no 
requirements on number of hits. To reduce backgrounds from inelastic 
collisions, LM vetoes (no hits in either the LM counters on the proton 
and antiproton sides) were included as part of the elastic triggers. The 
timing scintillator in each detector is only used for providing time 
information for off-line analysis and is not part of the triggers.

\section{Elastic Event Selection}
\label{sec:selection}
 An initial data sample is obtained by requiring events to satisfy one of the 
 elastic triggers. The $p$ and \pbar\ hit coordinates are measured in the 
FPD system using the fibers information and then used to select the 
sample of elastic scattering events. We  align the
detectors with respect to the beam and then use the beam transport 
matrices ~\cite{BD2} (which are functions of the currents of the magnets located between the IP and the FPD detectors and correlate the $x$, $y$ coordinates and scattering angle of a particle at two specific $z$ locations) to 
reconstruct the paths of  protons and antiprotons through this
region of the Tevatron. Next, background subtraction
 and efficiency corrections are performed.  We use a Monte Carlo (MC) 
event generator to apply corrections for
acceptance, detector resolution, beam divergence and IP size effects.
 
\subsection{Data Sample}
\label{datasample}
 The data for this analysis were collected with dedicated beam conditions  designed to
facilitate the positioning of the FPD Roman pots as close to the beam axis as
possible. The Tevatron injection tune with 
the betatron function of $\beta^{\ast}=1.6$ m at the D0 IP was used
instead of the standard $\beta^{\ast}=0.35$ m lattice~\cite{BD}. Additionally, only one proton bunch and
 one antiproton bunch were present in the Tevatron.
Scraping in the vertical and horizontal planes to remove the halo tails of the
bunches was performed and the electrostatic separators were turned off before
 initiating collisions.
The initial instantaneous luminosity was
about $0.5 \times 10^{30}$ cm$^{-2}$s$^{-1}$ with a lifetime of about 30 hours, corresponding to a mean number of \mbox{$\approx$ 0.8} interactions per bunch crossing. The recorded luminosity is about \mbox{$\cal{L}$ = 31 nb$^{-1}$},  which corresponds to the sum of two data sets used in
 this analysis, each with different detector positions with respect to the  beams. One set of data 
was taken with the closest detector position reaching about 4 mm with respect to the beam (data set 1),  and the other 
data set was taken with detectors about 1 mm
closer to the beam (data set 2). For the given instantaneous luminosity and the conditions of this store, about 33$\%$ of the elastic events are expected to be produced together with an inelastic collision in the same bunch crossing.

Approximately 20 million events were
recorded using a special trigger list optimized for diffractive
physics, including triggers for elastic, single diffractive, and
double pomeron~\cite{FPD1} exchange. This analysis uses elastic triggers, which make 
up about 10$\%$ of the total data collected. Independent triggers were used
to determine efficiencies.

\subsection{Hit Reconstruction}
\label{hit_reco}

We reconstruct the proton and antiproton hits using fibers with an associated signal above the 
same threshold as used in the FPD trigger system, that is tuned to accept 
hits from a minimum ionizing particle. For each fiber in an un-primed layer above threshold, we verify 
 if there is an adjacent fiber  in the primed layer in order to define
a ``fine'' fiber segment through which the particle traversed the plane.
Events with more than four fibers firing in a plane are typically not
due to single particles and are rejected (more than 99$\%$ of the elastic triggered events survive this condition). We require at least two
out of the three UU$^{\prime}$, VV$^{\prime}$, and XX$^{\prime}$ fine segments to be reconstructed in
each detector, with their
intersection yielding the transverse coordinates of the hit. Since noise in some of the fibers could produce fake hits, in the case of 
 more than one reconstructed hit in a detector (which happens in about 10$\%$ of the events), we weight the hits by the
sum of the analog-to-digital converter (ADC) pulse height of all the fibers that contributed to each hit and retain only the hit with the highest 
weight. The correction for the selection efficiency, discussed  in Sec.~\ref{sec:effi}, accounts for
elastic events  discarded by this requirement. 

 Utilizing events with fine segments in U, V, and X, we are able to
estimate the offset in position and resolution within each detector by taking the
difference between the $x$ coordinate obtained from UV fiber intersection
and a similar measurement using the X fiber plane. 
Most of the detectors used in this analysis have a resolution of about \mbox{150 $\mu$m} with the exception of some of the A2 detectors that had several inactive scintillating fibers, degrading their resolution to about \mbox{250 $\mu$m}.

\subsection{Selection of Candidate Elastic Events}

\begin{figure}
\centerline{\includegraphics[scale=0.5]{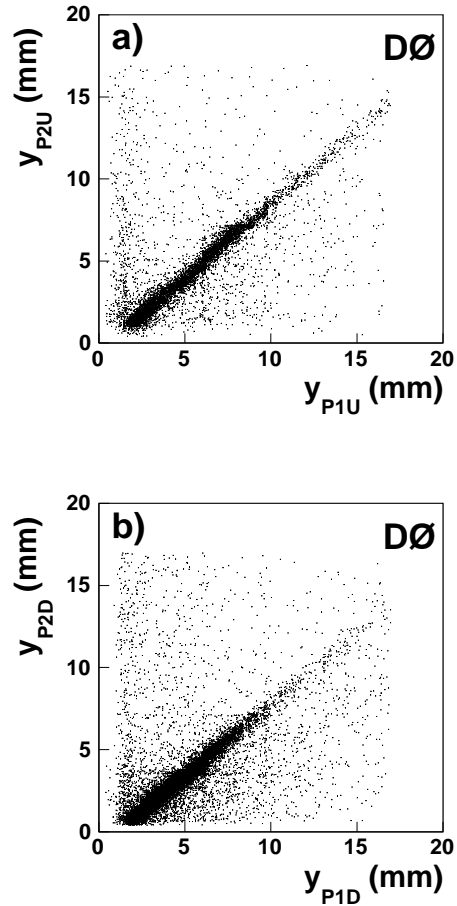}}
\caption{Comparison of detector $y$ coordinates in the spectrometers a) P$_{\text{U}}$   and b) P$_{\text{D}}$.} \label{plt:yyspect}
\end{figure}

Elastic collisions produce one of the four possible hit configurations in the FPD: A$_{\text{U}}$P$_{\text{D}}$, A$_{\text{D}}$P$_{\text{U}}$, A$_{\text{I}}$P$_{\text{O}}$, A$_{\text{O}}$P$_{\text{I}}$. Since  the inner and outer detectors are farther from the beam than the U and D detectors, they have poorer acceptance for elastic events and are only used for alignment purposes (see Sec.~\ref{sec:al}).
If we compare the hit coordinates reconstructed by each of the two detectors of one spectrometer, we observe a
 correlation band from  particles going through the
spectrometer, but also observe some uncorrelated background hits (see Fig.~\ref{plt:yyspect}).
We require hits to lie within  $\pm$3$\sigma$ ($\sigma \approx$ 220 $\mu$m) from the center of the band. We determine the quantity $\sigma$ by
fitting a Gaussian distribution obtained by projecting the hits onto an axis perpendicular to  the correlation band.

We then investigate the correlation between the coordinates of the protons and
antiprotons in diagonally opposite spectrometers. In addition
to the expected correlation due to the  collinearity of elastic
events, some background contamination due to halo particles remains.
 This contribution is reduced through the use of
the FPD timing system, but  there is still some
residual background, partially due to inefficiency of the trigger
scintillators and to different acceptances of the spectrometers. A correction for the contribution 
from halo within the correlation band is discussed in Sec.~\ref{sec:bck}.

\begin{figure*}
 \includegraphics[scale=0.8]{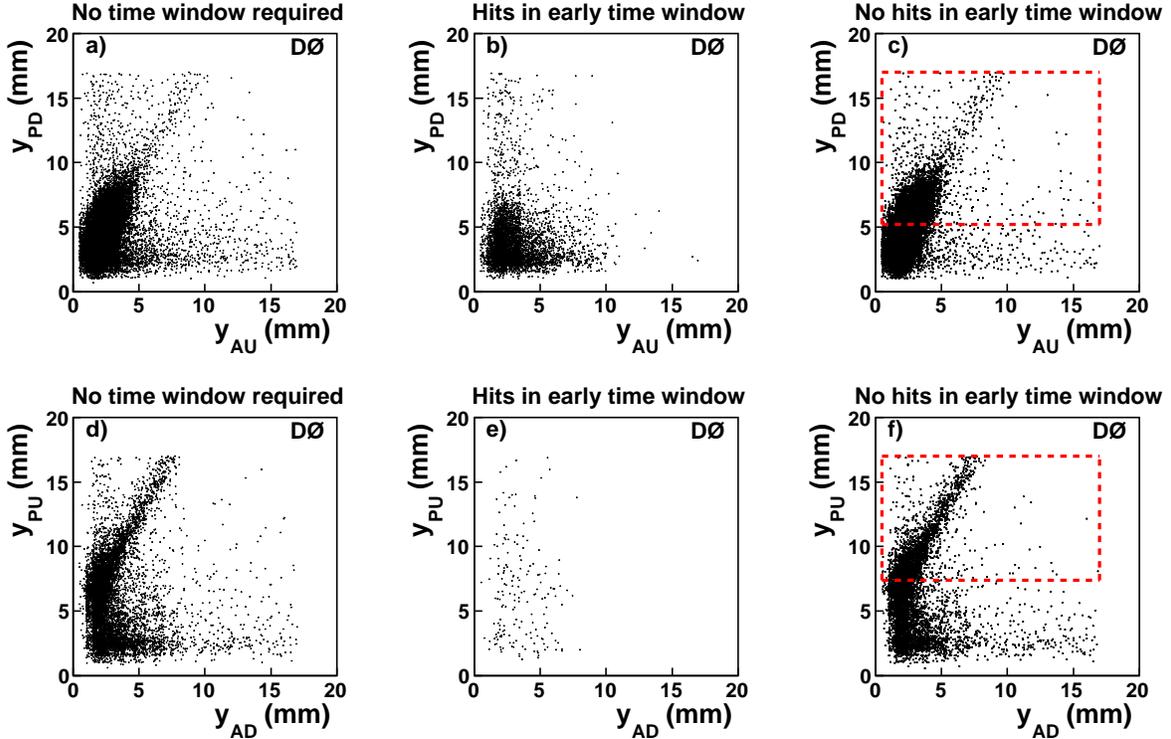} 
\caption{The correlation plots $y_{p}$ vs $y_{\ol p}$ for the first detectors in the spectrometers ($y$ coordinates measured in the local coordinate system 
shown in Fig.~\ref{plt:potpos2}) for A$_{\text{U}}$P$_{\text{D}}$ and 
A$_{\text{D}}$P$_{\text{U}}$. Plots a) and d) show the correlations 
without any timing requirement; b) and e) correspond to 
hits with a coincident tag in the corresponding scintillator 
early time windows; c) and f)  show  events with no hits within the early  
time window (most of the hits in this case are in the in-time window). The dashed lines correspond to the fiducial requirements applied.}
\label{plt:yyelas}
\end{figure*}

 For a specific set of $p$ and \pbar\ spectrometers in  an
elastic combination, we observe that elastic events with a proton
passing through a small region of the detectors in one spectrometer
will have  \pbar\ positions distributed over a similar but somewhat larger
region of the diagonally opposite spectrometer. Depending on the
location of the proton, the Gaussian distribution of the \pbar\
position may be truncated due to the finite detector size.  After
calculating the \pbar\ acceptance correction  as a function of the
coordinates of the proton
spectrometer, a fiducial cut is applied in the P$_1$ detectors such that only regions in these detectors which
require a correction of 2$\%$ or less are used. The 
region we select in the P$_1$ detectors is thus guaranteed to correspond to an
acceptance greater than 98$\%$ in the other three detectors. Figure~\ref{plt:yyelas} shows the $y_{p}$ vs $y_{\ol p}$ coordinate correlation plots ($y$ coordinate
 measured in the local coordinate system shown in Fig. \ref{plt:potpos2}), the 
tagging of the hits according to time-of-flight information, and the fiducial 
cuts applied (indicated by the dashed lines). We also check that there is no activity in the calorimeter for 
events in the elastic data sample. The contribution to the elastic sample from 
events that have a non-zero energy in the calorimeter is less 
than 0.1\% of the total number of selected events. Most of the elastic events produced in beam crossings with multiple $p$\pbar\ interactions are suppressed by the vetoes on the LM detectors.

\subsection{Forward Proton Detector Alignment} 
\label{sec:al}
In order to reconstruct the tracks from the protons and antiprotons, we 
first align the detectors with
respect to the beam and then determine
the position and angle of a particle at the IP  from the
measured hit positions in the FPD detectors after application of the
Tevatron transport matrices.

\begin{figure}
\centerline{\includegraphics[scale=0.45]{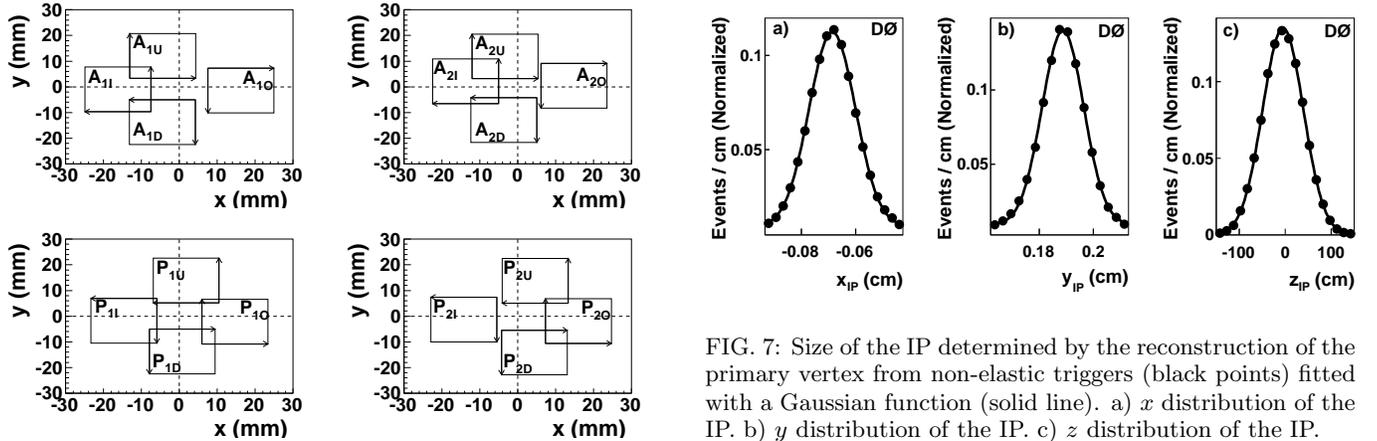}} 
\caption{Detector positions with respect to beam center (dashed lines) for 
the data set corresponding to the closest pot insertion to the beam. The arrows
indicate the local coordinate system in each detector.}
\label{plt:potpos2}
\end{figure}

The location of the detectors with
respect to the beam is determined  using  a sample of tracks that pass
 through one vertical and one horizontal detector at the same Roman pot station, allowing  
determination of
the relative alignment of the detectors. We use elastic events to align one horizontal detector that did not have any overlap with the other detectors at the same $z$ location.
 We define the center of a
 vertical and a horizontal pot at each Roman pot station and then
measure the positions of all four detectors with respect to this
reference system. Due to the beam optics the $x$ ($y$) hit coordinate distributions in the vertical proton (horizontal antiproton) detectors are narrower than in the vertical antiproton (horizontal proton) detectors. Because of this effect, the offset of the beam  in $x$ for the
vertical proton detectors (and the offset of the beam  in $y$ for the
horizontal antiproton detectors) is determined by fitting a Gaussian distribution to the $x$ and
$y$ distributions in these detectors. All other beam offsets are obtained given the fact
that the reconstructed IP offset and scattering angle is the same for both the 
proton and antiproton in an elastic event. We take the average over all four
elastic combinations to determine the offset of the beam at each Roman pot
station. 
Figure~\ref{plt:potpos2} shows the position of each detector obtained with the alignment procedure described
above for the data set corresponding to the detector configuration
with Roman pots located closest to the beam. The coordinates are plotted with respect to the beam coordinate system (dashed lines shown in the figure). For the data collected with the Roman pots retracted further away  from the beam line,
 we use positional difference information obtained from the pot motion system
  added to the previously
determined aligned   positions. 
The uncertainty of the location of the 
detectors after the 
alignment procedure is estimated to be about 200 $\mu$m.

\subsection{Reconstruction of $\boldsymbol{p/\bar{p}}$ tracks and $\boldsymbol{|t|}$ measurement}

  The Tevatron transport matrices are unique for these data due 
 to the use of the injection lattice. The beam radius (at 1$\sigma$ level) ranges from 0.4 mm to 0.8 mm at the different FPD detector
 locations while the beam divergence is about \mbox{44 $\mu$rad}. The change in the proton or antiproton divergence  caused by the quadrupoles system is typically of the order of 1 mrad for the smallest $|t|$ values considered in this measurement.

\begin{figure}
\includegraphics[scale=0.45]{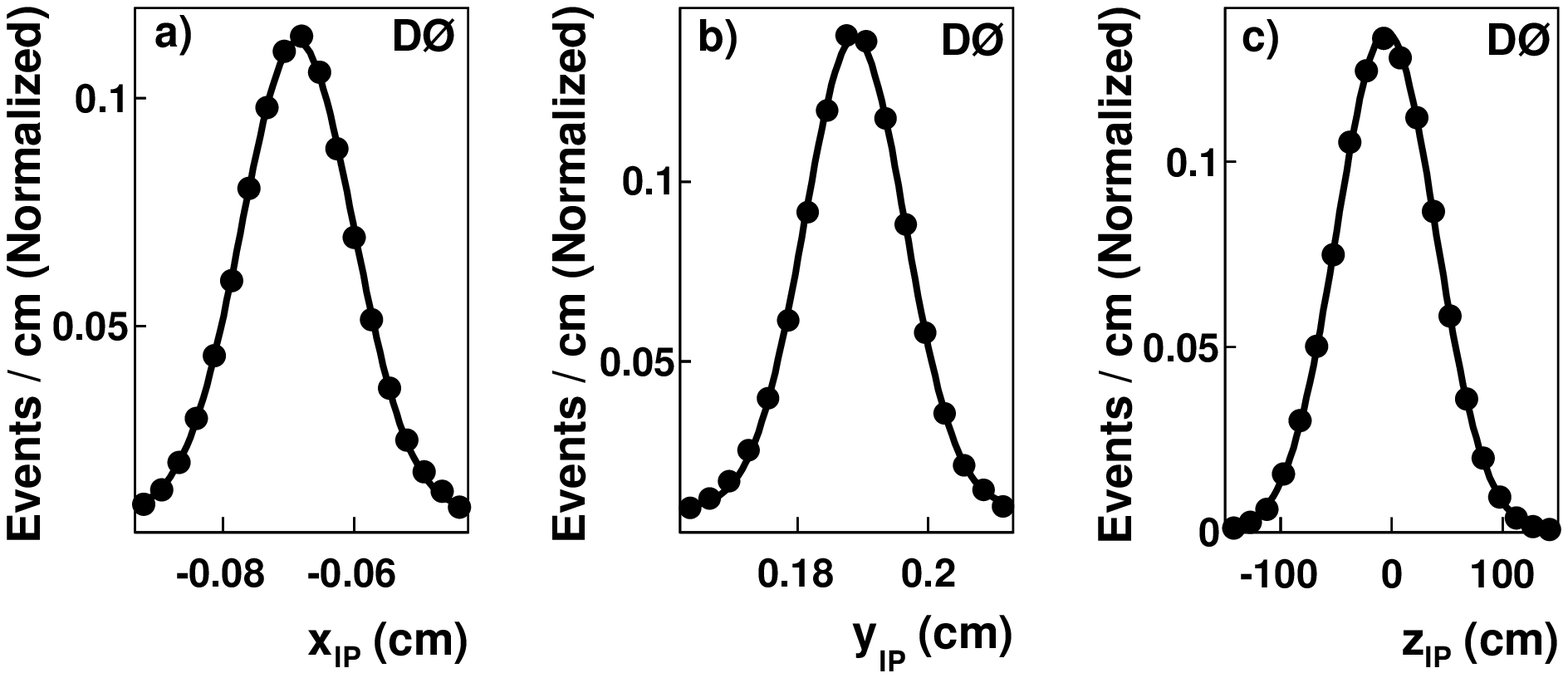}
\caption{Size of the IP determined by the reconstruction of the primary vertex from non-elastic 
triggers (black points) fitted with a Gaussian function (solid line). a) $x$ distribution of the IP. b) $y$ distribution of the IP. c) $z$ distribution of the IP.} 
\label{plt:beam_spot}
\end{figure}

 The size of the IP  in $x$, $y$ and $z$ is typically determined by 
reconstructing the primary vertex using the tracking detectors \cite{FPD}. Given the absence of 
central tracks in elastic events, data from non-elastic  triggers, simultaneously collected with 
the elastic scattering events used in this analysis, are used to obtain the $x$, $y$ and $z$ distributions 
of the  IP. The size of the IP in each direction is obtained by fitting a Gaussian function to each 
distribution. The measured Gaussian width of the IP distribution is about 100 $\mu$m in the transverse plane and about 45 cm 
along the $z$ axis (see Fig.~\ref{plt:beam_spot}). The offsets observed in $x$ and $y$  from the primary 
vertex reconstruction are due to the fact that the center of the tracker is shifted  with respect to 
the beam line. We use the reconstructed elastic events to determine these offsets, as described below.

\begin{figure}
\includegraphics[scale=0.45]{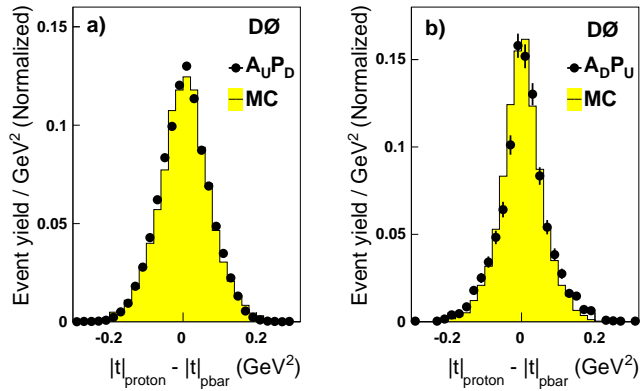}
\caption{Difference in the reconstructed $|t|$ between $p$ and \pbar\
 vertical spectrometers. The MC (solid line) expected distribution is compared to data (solid points). a)A$_{\text{U}}$P$_{\text{D}}$ combination. b) A$_{\text{D}}$P$_{\text{U}}$  combination. The number of entries in each distribution is normalized to unity. } \label{plt:deltat}
\end{figure}

 To tag an elastic event, we  reconstruct the coordinates of the hits in the two proton and two
antiproton detectors. The difference of coordinates between a proton and antiproton 
detector yields a Gaussian distribution with a width related to the
fiber offset and resolution of the two detectors, the IP size, and the beam
divergence. Given the previously estimated detector resolutions
and IP size, these distributions can be used to estimate the beam
divergence. The values obtained are similar (\mbox{40 $\pm$ 5 $\mu$rad}) to those estimated using 
the injection tune lattice parameters.
With the coordinates of the two detectors in each spectrometer we determine the
 offsets of the IP ($x_0$, $y_0$) and the horizontal and vertical scattering angles ($\theta_x$, $\theta_y$ respectively)  by using 
the following equations: 
\begin{eqnarray}
x_{i} & = & M_{x,i}x_{0} + L_{x,i}\theta_{x} \\
y_{i} & = & M_{y,i}y_{0} + L_{y,i}\theta_{y}  \nonumber
\end{eqnarray}
where $x_i$, $y_i$ are the hit coordinates  in a detector $i$ and $M_{x,i}$, $L_{x,i}$ ($M_{y,i}$, $L_{y,i}$) are 
the transport matrix elements in the horizontal (vertical) axis for that detector location. We verify that the distributions of the difference 
of the IP offsets and scattering angles obtained from the proton spectrometer and the antiproton spectrometer for every event are centered around zero. 
Since the values of $\theta$ obtained are of the order of milliradians, the four-momentum-transfer squared, $t$, can be approximated as
\begin{eqnarray}
t & = -p^2(\theta_x^2 + \theta_y^2) 
\end{eqnarray}
where $p$ is the momentum of the scattered particle. Given that the momenta
of the elastically scattered proton and antiproton are the same as that of the incoming particles (980 GeV) and that the beam 
momentum spread is negligibly small (0.014$\%$)~\cite{TEV}, the uncertainty in $|t|$  is dominated by the measurement uncertainty of the
scattering angle. Figure~\ref{plt:deltat} shows the difference in
the reconstructed $|t|$ from the \pbar\ and $p$  vertical
spectrometers, $\Delta |t|=|t|_{\ol p}-|t|_{p}$. The non-zero standard deviation of
$\Delta |t|$, $\sigma_{\Delta t}$, is due to the resolution on $|t|_{p}$ and $|t|_{\ol p}$.
The average of  $|t|_{p}$ and $|t|_{\ol p}$ ($|t|_{\text{ave}}=(|t|_{p}+
|t|_{\ol p})/2$) has a resolution of approximately
$\sigma_{\Delta |t|}/2$. The $|t|$ bin size is chosen as the largest of the observed $\sigma_{\Delta |t|}$ for the two detector combinations A$_{\text{U}}$P$_{\text{D}}$ and A$_{\text{D}}$P$_{\text{U}}$, where the difference in $\sigma_{\Delta |t|}$ results from different detector resolutions.  We also study the $|t|$
resolution as a function of $|t|$ and observe a gradual increase with
$|t|$, from \mbox{0.02 GeV$^{2}$} to \mbox{0.04 GeV$^{2}$}, which is reflected in the bin widths.

\subsection{\label{sec:bck}Background Subtraction} 

 The primary source 
of background in the selected sample is due to beam halo,
consisting of either   a  
halo proton and a halo antiproton  in the same bunch crossing or a halo
 particle
combined with a single diffractive event. A halo particle passing through the proton detectors in the time window for the protons which have undergone elastic scattering usually passes through the diagonally opposite antiproton detector at an earlier time (and vice versa). Therefore, the time information  can be used to veto events with early time
hits, consistent with halo protons and antiprotons in the elastic
sample. The veto is not 100$\%$ efficient due to a combination of
scintillator efficiency and  positioning of the
detectors with respect to the beam (a closer detector position both
increases a detector's signal acceptance and its ability to reject halo). Consequently, it is necessary
to subtract the remaining background. We consider an event to be 
caused by \pbar\ ($p$) halo if one or both of the two $p$ (\pbar) detectors of an 
elastic combination have hits in their early time interval. We require events to have no activity in either the $\ol p_{\text{halo}}$ or the $p_{\text{halo}}$ timing
window. We  select background samples by requiring hits consistent with $p$ halo and \pbar\ halo
simultaneously. First, we verify that outside the elastic correlation band between the coordinates of a proton and antiproton detector the signal tagged events have the same $|t|$ dependence as the background tagged ones (see Fig.~\ref{plt:yyelas}). Next, assuming that signal and background tagged events also have the same $|t|$ dependence inside the correlation band, we use the ratio 
of these two distributions and use it to estimate the percentage of 
background events inside the signal tagged correlation band. This background is subtracted from all 
events inside the correlation band to obtain $dN$/$dt$ as a function of $|t|$.  The amount of background subtracted inside the elastic correlation
band varies from 1$\%$ at low $|t|$  to 5$\%$ at high $|t|$. The absolute uncertainty of the background, which is propagated as a statistical uncertainty to $d\sigma$/$dt$,  varies from 0.3\% at low $|t|$  to 5.0\% at high $|t|$.
As a cross check, we vary the detector band cuts from 3.0$\sigma$ to 3.5$\sigma$ and to 6.0$\sigma$, to allow more background, and 
obtain similar $dN$/$dt$ results after applying the same background subtraction procedure (within 1$\%$).

\subsection{Monte Carlo Simulation of Elastic Events}
 \label{mc} 

We have developed a MC generator interfaced with the
Tevatron transport matrices to generate elastic events. 
The MC allows us to study the
geometrical acceptance of the detectors, resolution of the position 
measurement,
alignment, and effects of the beam size and beam divergence at the
IP. The generation of events is based on an Ansatz
function that we obtain by fitting the $dN$/$dt$
distribution of the data. We study the acceptance and bin migration effects
 using samples generated with a wide range of different $dN$/$dt$
Ansatz distributions. The variations in the corrections are included as 
systematic uncertainties.

\begin{figure}
\includegraphics[scale=0.45]{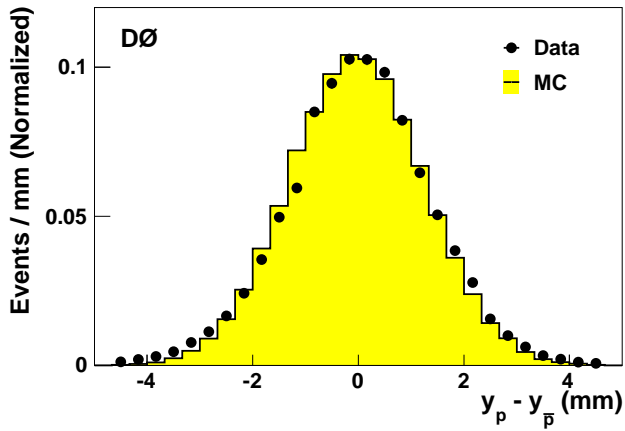}
\caption{Comparison of MC (solid line) and data (black points) for the coordinate difference $y_p$ - $y_{\ol p}$  for the elastic configuration A$_{\text{1,U}}$P$_{\text{1,D}}$ for the data set with detectors closer to the beam.}
\label{plt:yymc}
\end{figure}

The positions 
of hits in each detector are transformed into fiber hit information, and the
reconstruction then proceeds using these hits, following the same
procedure as with the data.
The reconstructed correlation patterns in MC are in good
agreement with those observed in data. In addition, the MC also
predicts the widths of the different correlations as  shown in Fig.~\ref{plt:deltat} and Fig.~\ref{plt:yymc}  .

\subsection{\label{sec:acc}Acceptance and Bin Migration Correction}

\begin{figure}
\includegraphics[scale=0.45]{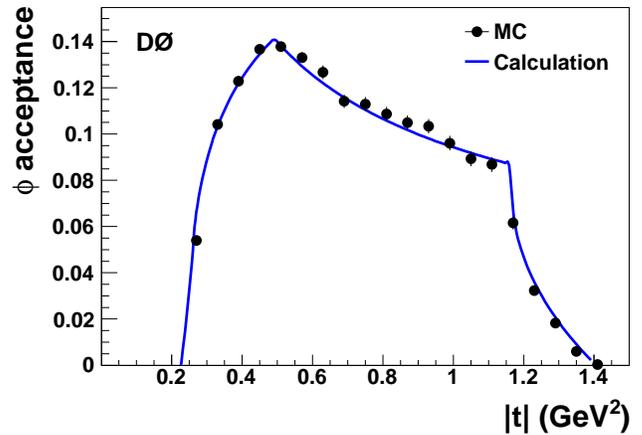}
\caption{Azimuthal acceptance of the FPD  in data, after
fiducial cuts, for the closest detector position of the A$_{\text{U}}$P$_{\text{D}}$ elastic combination. The 
points correspond to MC, the solid line corresponds to the calculation of the $\phi$ acceptance 
from the detector positions and the fiducial area. }
\label{plt:phiacc}
\end{figure}

Due to accelerator optics, protons and antiprotons of a particular $|t|$ are transported to an 
elliptical region in the $x-y$ plane at each detector location. Each detector has a different coverage in  
the azimuthal angle $\phi$  and therefore a different coverage of the ellipses of 
constant $|t|$. The $\phi$ acceptance correction 
accounts for the fraction of the $|t|$ ellipse for each $|t|$ bin that
is not covered by the fiducial area used in each detector. This correction  depends only on 
the detector location with respect to the particle
beam and on the geometry of the fiducial area used in the detector for selecting the 
events. Figure~\ref{plt:phiacc} shows the $\phi$ acceptance calculated 
as a function of $|t|$ for data at the closest detector
position with respect to the beam and after the fiducial cuts. We obtain similar results for the
$\phi$ acceptance using the MC described in the previous
section before adding the effects of beam divergence, IP size and detector resolution. The uncertainty in the $\phi$ acceptance correction, which comes from the size of the MC sample used, is less than 0.1\% at low $|t|$ values 
and less than 1.0\% at high $|t|$ values.
To estimate the correction for bin migration effects that is applied in data to obtain the $d\sigma$/$dt$ distribution, we include the measured values of beam divergence, IP size, detector resolution and pot position uncertainties  in the MC generator and compare to a same size MC sample with no smearing effects.

\subsection{\label{sec:effi}Selection and Trigger Efficiencies} 

 We simultaneously determine
the effects of selection and trigger efficiencies for each of the
four detectors corresponding to an  elastic configuration. To determine the
efficiency of a particular detector, we use an independent trigger
which does not include that detector. 
 We obtain the $dN$/$dt$ distributions with an elastic track reconstructed in the other three detectors and for all four detectors, with the trigger conditions satisfied. The ratio of
the two distributions is used to extract the efficiency of the detector
 as a function of $|t|$ (where $|t|$ is reconstructed from the
coordinates of the opposite side spectrometer).
 To avoid any additional effect from 
detector acceptance, we use only  hits in the other three detectors in
a region determined to have full geometrical acceptance in the detector of interest. We
repeat a similar procedure for each of the four detectors in every
elastic combination and multiply the efficiencies of the four
detectors to determine the final efficiency correction.
Typical selection and trigger efficiencies are in the range of 50$\%$ to 70$\%$
depending on the detector and trigger requirements.
Additionally, we  make a correction for the veto in the LM which was part of the elastic trigger. This veto filters out elastic 
events that are produced in coincidence with an inelastic collision in the same bunch crossing 
(pileup). To make this correction, we use a trigger based on hits in the proton spectrometers with 
no LM requirement and determine the fraction of candidate elastic events reconstructed in coincidence 
with the LM. We find that about 27$\%$ (20$\%$) of elastic events for 
data set 1 (2) as defined in Sec.~\ref{datasample} were removed by the LM veto.

\subsection{\label{sec:lum}Luminosity }
Since elastic data were collected with  Tevatron
conditions modified with respect to standard operations of the D0 experiment, we compare the number of 
inclusive jet events obtained from our data to the number of jet events used for the inclusive cross section measurement discussed in~\cite{INCLUSIVE} to determine the integrated luminosity.
The corresponding integrated luminosity for 
data set 1 (2) is  \mbox{18.3 nb$^{-1}$} (\mbox{12.6 nb$^{-1}$}), with an uncertainty of 13$\%$. We add in quadrature the uncertainty in the standard luminosity determination (6.1$\%$~\cite{lumi}) and obtain an
overall normalization uncertainty of 14.4$\%$.

\section{\label{sec:sys}Systematic Uncertainties }

 The major $|t|$-dependent contributions to the systematic uncertainties in the measurement of
$d\sigma$/$dt$ are due to detector efficiencies,
beam divergence, detector positions, and the choice of the Ansatz function.  The luminosity 
measurement contributes to the overall normalization uncertainty. The Tevatron beam transport matrices are known
with high precision (within 0.1$\%$) and therefore produce a  small uncertainty
in our results compared to the other sources. We take the uncertainty in the position of the pot from the beam center as an extra ``smearing'' factor for the hit coordinates in the MC. Since the efficiencies vary with $|t|$,  we fit either
a polynomial or an exponential function to each trigger efficiency
and propagate the uncertainties in the fit parameters to
$d\sigma$/$dt$, using the
covariance matrix of the fit. For the beam divergence term, we  vary the beam divergence by $\pm$5 $\mu$rad in the MC, and we
propagate the change in the acceptance correction bin-by-bin  to $d\sigma$/$dt$. We also consider 26
possible variations of the Ansatz function used in the MC to account for 
 the uncertainties in the logarithmic slopes before and after the kink and also
for the $|t|$ value where the kink is observed.

\section{\label{sec:res}Results }
\begin{figure}
\includegraphics[scale=0.44]{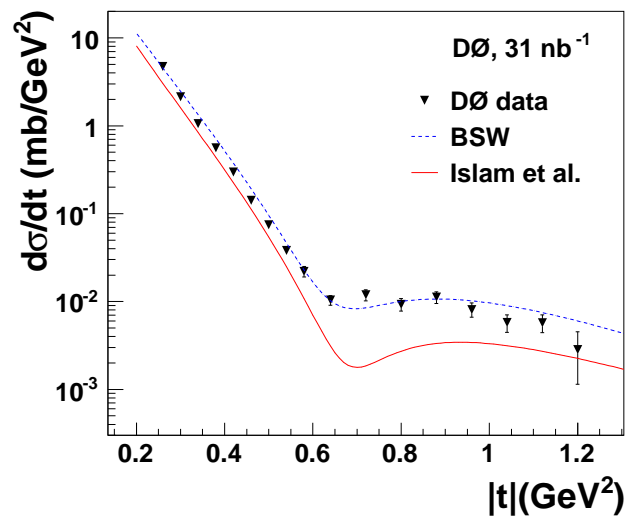}
\caption{The measured $d\sigma$/$dt$ differential cross section. The normalization uncertainty of 14.4$\%$ is not shown. The uncertainties are obtained by adding in quadrature statistical and systematic uncertainties. The predictions of BSW (\cite{soffer}) and Islam et al. (\cite{islam}) are compared to the data.} \label{plt:B}
\end{figure}

In total, we have four independent 
measurements of $d\sigma$/$dt$ that come from the two elastic
combinations (A$_{\text{U}}$P$_{\text{D}}$ and A$_{\text{D}}$P$_{\text{U}}$) and two data sets, which agree with each other within the uncertainties. We combine the
four measurements using a bin-by-bin weighted average. 
The resulting values for the $d\sigma$/$dt$ distribution, together with their total uncertainties, are listed in Table~\ref{tbl:dsdt} and shown in Fig.~\ref{plt:B}. The uncertainties are 
the total experimental uncertainties excluding the 14.4$\%$ normalization uncertainty. The $|t|$ bin centers are determined using the prescription described in~\cite{wyatt}, however, the values found are very close to the middle of the bin.
Two phenomenological 
model predictions for $\sqrt{s}=1.96$ TeV (BSW \cite{soffer}, Islam et al.~\cite{islam})  are 
also shown. The BSW model shows a good description of the data in shape and normalization and is able to reproduce the kink within experimental uncertainties.

\begin{table}
\caption{\label{tbl:dsdt} The $d\sigma$/$dt$ differential cross section. The statistical and systematic uncertainties are added
in quadrature. The luminosity uncertainty of
14.4$\%$ is not included.}
\hspace*{0.5cm}
\begin{ruledtabular}
\begin{tabular}{cc}
 $|t|$ (GeV$^{2}$)&$d\sigma$/$d|t|$ (mb/GeV$^{2}$)  \\ \hline
  0.26 & (47.3 $\pm$ 2.6)  $\times$ 10$^{-1}$  \\
  0.30 & (21.4 $\pm$ 0.72) $\times$ 10$^{-1}$  \\
  0.34 & (10.6 $\pm$ 0.37) $\times$ 10$^{-1}$  \\
  0.38 & (5.64 $\pm$ 0.22) $\times$ 10$^{-1}$  \\
  0.42 & (2.98 $\pm$ 0.14) $\times$ 10$^{-1}$  \\
  0.46 & (14.2 $\pm$ 0.83) $\times$ 10$^{-2}$  \\
  0.50 & (7.46 $\pm$ 0.51) $\times$ 10$^{-2}$  \\
  0.54 & (3.81 $\pm$ 0.36) $\times$ 10$^{-2}$  \\
  0.58 & (2.20 $\pm$ 0.30) $\times$ 10$^{-2}$  \\
  0.64 & (1.04 $\pm$ 0.13) $\times$ 10$^{-2}$  \\
  0.72 & (1.19 $\pm$ 0.17) $\times$ 10$^{-2}$  \\
  0.80 & (9.28 $\pm$ 1.5)  $\times$ 10$^{-3}$  \\
  0.88 & (11.2 $\pm$ 1.7)  $\times$ 10$^{-3}$  \\
  0.96 & (8.11 $\pm$ 1.5)  $\times$ 10$^{-3}$  \\
  1.04 & (5.77 $\pm$ 1.3)  $\times$ 10$^{-3}$  \\
  1.12 & (5.73 $\pm$ 1.3)  $\times$ 10$^{-3}$  \\
  1.20 & (2.84 $\pm$ 1.7)  $\times$ 10$^{-3}$  \\
\end{tabular}
\end{ruledtabular}
\end{table}

The $|t|$ range covered by our measurement is \mbox{$0.26 <|t|< 1.2$ GeV$^{2}$}. We observe a change in the logarithmic slope of 
the $d\sigma$/$dt$ distribution at \mbox{$|t| \approx 0.6$ GeV$^{2}$}. 
A fit to the $d\sigma$/$dt$ distribution in
the range \mbox{$0.26<|t|<0.6$}  with an exponential function of
the form $Ae^{-b|t|}$ yields a logarithmic slope parameter of  
\mbox{$ b = 16.86$ $\pm$ $0.10$ (stat) $\pm$ $0.20$ (syst) GeV$^{-2}$} (\mbox{$\chi^2=6.63$ for 7 degrees of freedom}). 

\begin{figure}
\includegraphics[scale=0.44]{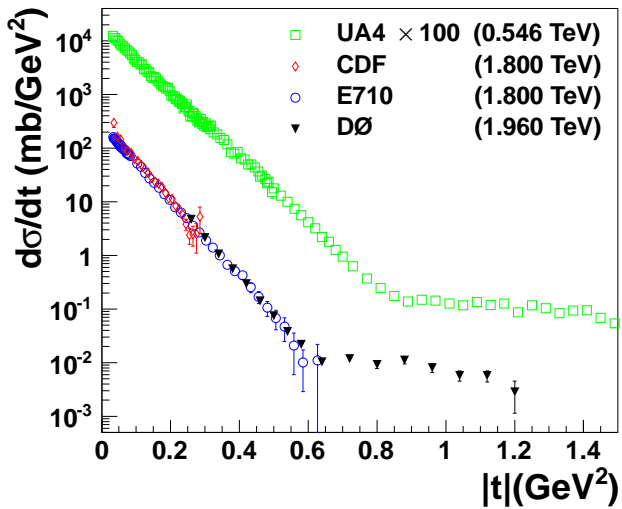}
\caption{The $d\sigma$/$dt$ differential cross section measured by the D0 Collaboration and compared to the CDF and E710 measurements at $\sqrt{s}=1.8$ TeV, and to the UA4 measurement at $\sqrt{s}=0.546$ TeV (scaled by a factor of 100). A normalization uncertainty of
14.4$\%$ on the D0 measurement is not shown.}
\label{plt:tfinal}
\end{figure}

Figure~\ref{plt:tfinal} shows a comparison of our results to those obtained 
at $\sqrt{s}=1.8$ TeV by  the CDF and E710
Tevatron Collaborations~\cite{CDF:B,E710:larget}, and also to that of the UA4 Collaboration at $\sqrt{s} = 0.546\  \mathrm{TeV}$~\cite{ua4:dsdt}. Our measurement of  the 
slope parameter agrees within uncertainties with previous measurements by the CDF ($b = 16.98 \pm 0.25$ GeV$^{-2}$) and E710 ($b = 16.30 \pm 0.30$ GeV$^{-2}$) Collaborations. A comparison of the shape of our measured $d\sigma$/$dt$ to UA4 measurement shows that the kink in $d\sigma$/$dt$ moves towards lower $|t|$ values as the energy is increased, but, as in the UA4 data, we do not see a distinct minimum as observed in $pp$ elastic interactions (\cite{totem}).

In summary, we have presented  the first measurement of $d\sigma (p \ol p \rightarrow p \ol p)$/$dt$ as a function of $|t|$ at $\sqrt{s}=1.96$ TeV. Our measurement extends the $|t|$ and $\sqrt{s}$ range previously studied  and shows a 
change in the $|t|$ dependence, consistent with the features expected for the transition between two diffractive regimes.

\section{\label{sec:ack}ACKNOWLEDGEMENTS }
%
We thank the Fermilab Beams Division for designing and providing the special beam conditions for the data taking. In particular we thank N. Mokhov, S. Drozhdin, 
M. Martens and A. Valishev for their important contributions to the FPD. We also thank J. Soffer and M. Islam for useful discussions.  

We thank the staffs at Fermilab and collaborating institutions,
and acknowledge support from the
DOE and NSF (USA);
CEA and CNRS/IN2P3 (France);
MON, Rosatom and RFBR (Russia);
CNPq, FAPERJ, FAPESP and FUNDUNESP (Brazil);
DAE and DST (India);
Colciencias (Colombia);
CONACyT (Mexico);
NRF (Korea);
FOM (The Netherlands);
STFC and the Royal Society (United Kingdom);
MSMT and GACR (Czech Republic);
BMBF and DFG (Germany);
SFI (Ireland);
The Swedish Research Council (Sweden);
and
CAS and CNSF (China).
%


\begin{thebibliography}{99}
\bibitem{tt} $t=(p_f - p_i)^{2} $ where $p_i$ and $p_f$ are the initial and final particles four-momenta, respectively.
\bibitem{eta} The pseudorapidity is defined as $\eta=-\ln[\tan(\theta/2)]$, where $\theta$ is the polar angle with respect to the proton beam direction.
\end{thebibliography}
\end{document}